\documentclass[%
 reprint,
 amsmath,amssymb,
 aps,
 twocolumn,
]{revtex4}

\usepackage{graphicx}% Include figure files
\usepackage{dcolumn}% Align table columns on decimal point
\usepackage{bm}% bold math

\input{colordvi.tex}

\newcommand\aj{{AJ}}% Astronomical Journal
\newcommand\araa{{ARA\&A}} % Annual Review of Astron and Astrophys
\newcommand\apjl{{ApJ}}% Astrophysical Journal, Letters
\newcommand\apjs{{ApJS}}% Astrophysical Journal, Supplement
\newcommand\aap{{A\&A}}% Astronomy and Astrophysics
\newcommand\mnras{{MNRAS}}% Monthly Notices of the RAS
\newcommand\nar{{New A Rev.}}% New Astronomy Review
\newcommand\pasj{{PASJ}}% Publications of the ASJ
\begin{document}

\preprint{APS/123-QED}

\title{Upper Limit on the Cosmological Gamma-ray Background}% Force line breaks with \\

\author{Yoshiyuki Inoue}
 \email{yinoue@slac.stanford.edu}
\affiliation{Kavli Institute for Particle Astrophysics and Cosmology, Department of Physics and SLAC National Accelerator Laboratory, Stanford University, Stanford, CA 94305}

\author{Kunihito Ioka}
 \email{kunihito.ioka@kek.jp}
\affiliation{Theory Center, Institute of Particle and Nuclear Studies, KEK, 1-1 Oho, Tsukuba 305-0801, Japan}
\affiliation{The Graduate University for Advanced Studies (Sokendai), 
Oho 1-1, Tsukuba 305-0801, Japan}

\date{\today}% It is always \today, today,

\begin{abstract}
We show that the current extragalactic gamma-ray background (EGB)  measurement below 100 GeV sets an upper limit on EGB itself at very high energy (VHE) above 100 GeV. The limit is conservative for the electromagnetic cascade emission from VHE EGB interacting with the cosmic microwave-to-optical background radiation not to exceed the current EGB measurement. The cascade component fits the measured VHE EGB spectrum rather well. However, once we add the contribution from known source classes, the {\it Fermi} VHE EGB observation exceeds or even violates the limit, which is approximated as $E^2dN/dE<4.5\times10^{-5}(E/100\,{\rm GeV})^{-0.7}\ {\rm MeV/cm^2/s/sr}$. The upper limit above 100 GeV is useful in the future to probe the EGB origin and the new physics like axion-like particles and Lorentz-invariance violation.
\end{abstract}


\maketitle

\section{\label{sec:intro}Introduction}
The origin of the unresolved extragalactic diffuse gamma-ray background (EGB) radiation has been a big puzzle in astrophysics and astroparticle physics. The EGB was first discovered by the {\it SAS}--2 satellite \cite{fic78}. EGRET (Energetic Gamma-Ray Experiment Telescope) on board the Compton Gamma-Ray Observatory confirmed the EGB spectrum at 0.03-50 GeV \cite{sre98}. Recently,  LAT (Large Area Telescope) on board the Fermi Gamma-ray Space Telescope ({\it Fermi}) made a new measurement of the EGB spectrum from 0.2 to 100 GeV \cite{abd10_egrb}. The observed integrated EGB flux ($E>100$ MeV) is $1.03\times10^{-5}$ photons/cm$^2$/s/sr with a photon index of 2.41$\pm0.05$. This power-law spectrum extends up to 600 GeV based on the very recent preliminary EGB spectrum reported by the {\it Fermi} collaboration \cite{ack11_TeVPA}. 

EGB is composed of various unresolved gamma-ray sources. Point sources detected by EGRET and {\it Fermi} are guaranteed to contribute to EGB. Those are namely blazars \cite[e.g.][]{abd10_marco}, radio galaxies \cite{ino11}, starburst galaxies \cite[e.g.][]{ste11,ack12_stb}, high latitude pulsars \cite{fau10}, and gamma-ray bursts (GRBs) \cite{cas07}. It is expected that blazars, radio galaxies, and starburst galaxies explain $22.5\pm1.8$\% \cite{abd10_marco}, $25^{+38}_{-15}$\% \cite{ino11} and $4-23$\% \cite{ack12_stb} of the unresolved EGB, respectively. Other extragalactic sources have also been discussed as the origin of EGB, although they are still not detected in gamma-ray \cite[see][and references therein]{ino11}.

Very high energy (VHE; $\gtrsim30 {\rm GeV}$)  gamma-rays propagating through the universe experience absorption by the interaction with the extragalactic background light (EBL) via electron--positron pair production \cite[e.g.][]{fin10}. As discussed in \cite{ino11}, if the EGB radiation originates from cosmological sources, the EBL absorption signature should appear in the spectrum above $\sim$30 GeV. However, the measured EGB spectrum shows a single power-law up to 600 GeV \cite{abd10_egrb,ack11_TeVPA}. This may pose a serious problem for the current models.

Electron--positron pairs created by VHE gamma-rays with EBL scatter the cosmic microwave background (CMB) radiation via the inverse Compton (IC) scattering  and generate secondary gamma-ray emission component (the so-called cascade emission) in addition to the absorbed primary emission \cite[e.g.][]{fan04}. At redshift $z$, the scattered photon energy $E_{\gamma,c}$ appears at lower energy than the intrinsic photon energy $E_{\gamma,i}$, typically 
\begin{equation}
\label{eq:ecas}
E_{\gamma,c}\approx 0.8\,(1+z)\left(\frac{E_{\gamma,i}}{1\,{\rm TeV}}\right)^2 \ \ {\rm GeV}.
\end{equation}
The cascade component is also expected to contribute to EGB  \cite{cop97,mur12}. Recently Murase et al. (2012) \cite{mur12} constrained the cosmic energy density of gamma-rays using the cascade component contribution to the EGB from the {\it Fermi} measurement.

In this paper, we generalize the argument in a conservative way for arbitrary cosmological sources, and {\it set an upper limit on EGB by itself} with the new {\it Fermi} EGB data, in particular, on VHE EGB by requiring the cascade emission not to exceed the currently observed EGB below $\sim$30 GeV. Taking into account the guaranteed source's contributions, we find that the current EGB measurement already self-limits the VHE EGB fluxes above 100 GeV to PeV as
 \begin{equation}
E^2\frac{dN}{dE}<4.5\times10^{-5}
\left(\frac{E}{100\,{\rm GeV}}\right)^{-0.7}\  {\rm MeV/cm^2/s/sr}
\label{eq:limit_mo}
\end{equation}
for cosmological sources, which is {\it inconsistent} with the current EGB  measurement by {\it Fermi}. We further discuss the requirements for possible origins of VHE EGB. Hereafter we use the standard cosmology $(H_0,\Omega_M,\Omega_\Lambda)=(70.0\ {\rm km/s/Mpc},0.3,0.7)$.

\section{The EGB spectrum}
\label{sec:egb}

The EGB spectrum in the unit of ${\rm MeV^2/cm^2/s/sr/MeV}$ is calculated as 
\begin{eqnarray}
\nonumber
E^2\frac{dN}{dE}(E_{\rm obs}) &=&\frac{cE_{\rm obs}^2}{4\pi}\int_0^{z_{\rm max}}dz \left|\frac{dt}{dz}\right|(1+z)\\
&\times& \frac{dj}{dE_{\gamma}}[(1+z)E_{\rm obs},z]\exp[-\tau_{\gamma\gamma}(E_{\rm obs},z)], \nonumber \\
\end{eqnarray}
where $E_{\rm obs}$ is the observed photon energy, $c$ is the light speed, $t$ is the cosmic time, $\left|dt/dz\right|^{-1}=H_0(1+z)\sqrt{\Omega_M(1+z)^3+\Omega_\Lambda}$, and $\tau_{\gamma\gamma}(E_{\rm obs},z)$ is the gamma-ray opacity for $E_{\rm obs}$ from $z$. We assume $z_{\rm max}=5$, which does not affect our results. 
 
The comoving volume emissivity $dj/dE_{\gamma}(E_{\gamma},z)$ $[{\rm ph/s/MeV/cm^3}]$  is given by the intrinsic plus cascade emission, $j=j_{\rm int}+j_{\rm cas}$.

The intrinsic emission can be characterized by a few parameters,
\begin{eqnarray}
\label{eq:jint}
\frac{dj_{\rm int}}{dE_{\gamma}}(E_{\gamma},z)= 
\left\{\begin{array}{ll}
j_0 E_{\gamma}^{-\Gamma_{\rm ph}}(1+z)^{\beta_{\rm evo}}, &
E_{\gamma}\le E_{\rm max},\\
0, & 
E_{\gamma}>E_{\rm max},\\
\end{array}\right.
\end{eqnarray}
where $E_{\gamma}$ is the photon energy in the rest frame, the spectral shape is a power-law with a photon index $\Gamma_{\rm ph}$ and an cutoff at $E_{\rm max}$, the $z$-evolution is given by $\beta_{\rm evo}$, and $j_0$ is the normalization.

\begin{figure}[t]
\includegraphics[width=1\linewidth]{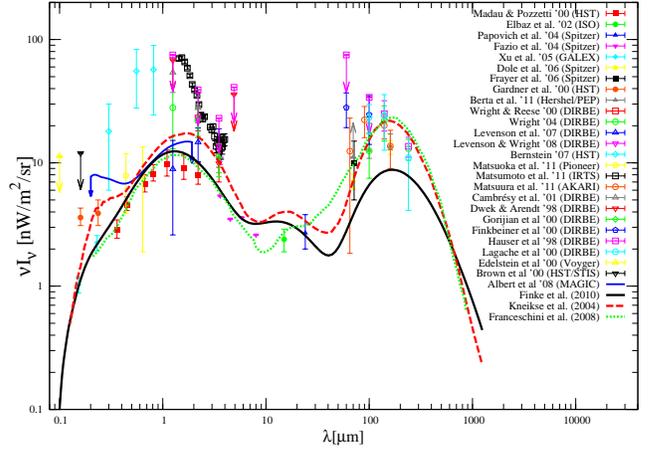}% Here is how to import EPS art
\caption{\label{fig:EBL} The EBL model \cite{fin10} used in this paper is shown by solid line. For comparison, other EBL models \cite{kne04} are shown as indicated in the figure. The integrated brightness of galaxies \cite{mad00} (minimum EBL; filled symbols)  and current measurements of the EBL \cite{wri04} (open symbols)  are shown as indicated in the figure. The upper limit from TeV gamma-ray observation by MAGIC \cite{alb08} is also shown by solid line with arrows. }
\end{figure}

 To give a conservative upper limit, we adopt the EBL model by \cite{fin10} as shown in Figure \ref{fig:EBL}. The EBL intensity by \cite{fin10} is close to the galaxy counts which are the lower limit of the EBL. As the EBL density becomes higher, the EGB upper limit gets tighter.

\section{Cascade Emissivity}

Following \cite{fan04,mur07}, we calculate the cascade emissivity ${dj_{\rm cas}}/{dE_\gamma}$ as:
\begin{equation}
\frac{dj_{\rm cas}}{dE_\gamma}(E_\gamma,z)=\int_{\gamma_{e,\rm min}}^{\gamma_{e,\rm max}} d\gamma_e\frac{dj_e}{d\gamma_e}\frac{d^2N_{\gamma_e,\epsilon}}{dtdE_{\gamma}}t_{IC}(z),
\label{eq:cascade}
\end{equation}
where $t_{IC}(z)$ is the energy-loss time of an electron with a Lorentz factor $\gamma_e$ and mass $m_e$ by the inverse Compton (IC) emission in the local rest frame,
\begin{equation}
t_{IC}(z)=\frac{3m_ec}{4\gamma_e\sigma_Tu_{\rm CMB}(z)}\approx7.7\times10^{13}\left(\frac{\gamma_e}{10^6}\right)^{-1} (1+z)^{-4}\  {\rm s},
\end{equation}
$\sigma_T$ is the Thomson scattering cross section, and $u_{\rm CMB}(z)$ is the CMB energy density at $z$. We consider the CMB photons only here, since the EBL energy density is two orders magnitudes lower than that of CMB.
$dj_e/d\gamma_e$ is the electron injection spectrum:
\begin{equation}
\frac{dj_e}{d\gamma_e} = 2 \frac{dE_{\gamma,i}}{d\gamma_e}\frac{dj_{\rm int}(E_{\gamma,i},z)}{dE_{\gamma,i}} 
\left[1 - e^{-\tau_{\gamma\gamma}(E_{\gamma,i}/(1+z),z)} \right],
\end{equation}
and $d^2N_{\gamma_e,\epsilon}/dtdE_{\gamma}$ is the scattered photon spectrum per unit time by the IC scattering:
\begin{equation}
\frac{d^2N_{\gamma_e,\epsilon}}{dtdE_{\gamma}}=\frac{3 \sigma_T c}{4 \gamma_e^2}\int d\epsilon \frac{1}{\epsilon}\frac{dn_{\rm CMB}}{d\epsilon}(\epsilon,z) f(x)
\end{equation}
with $f(x)=2x \ln(x)+x+1-2x^2$, ($0<x<1$) and $x = E_{\gamma}/4\gamma_e^2\epsilon$. Here, $E_{\gamma,i}=2\gamma_e m_e c^2$ is the energy of intrinsic photons and
$dn_{\rm CMB}/d\epsilon$ is the CMB photon density. The integration region over the Lorentz factor, $\gamma_e$, is $\gamma_{e,{\rm min}}<\gamma_e<\gamma_{e,{\rm max}}$, $\gamma_{e,{\rm max}}=E_{\rm max}/2m_ec^2$ and $\gamma_{e,{\rm min}}=(E_\gamma/\epsilon)^{1/2}/2$.  Since the cooling time $t_{\rm IC}$ is usually shorter than the comoving time, we assume that pairs generate photons at the pair creation site. Because of this fast cooling, the low energy photon spectrum below 100 MeV becomes $\Gamma_{\rm ph}=1.5$.  We do not take into account this spectral effect since it does not affect the VHE spectrum.

We iteratively calculate Eq.~(\ref{eq:cascade}) by substituting $(2 dj_{\rm cas}/d\gamma_e)(1-e^{-\tau_{\gamma\gamma}})$ for $dj_{e}/d\gamma_e$ in order to include IC scatterings due to  pairs from reabsorption of cascade photons \cite{mur07}. 

The intergalactic magnetic field (IGMF) effect is not important in this study. Although IGMF bends motion of created charged pairs and some fraction of beamed emission is lost, off-axis sources complement this loss. The synchrotron cooling is also not effective for pairs typically created outside galaxies.

\begin{figure}[t]
\includegraphics[width=1\linewidth]{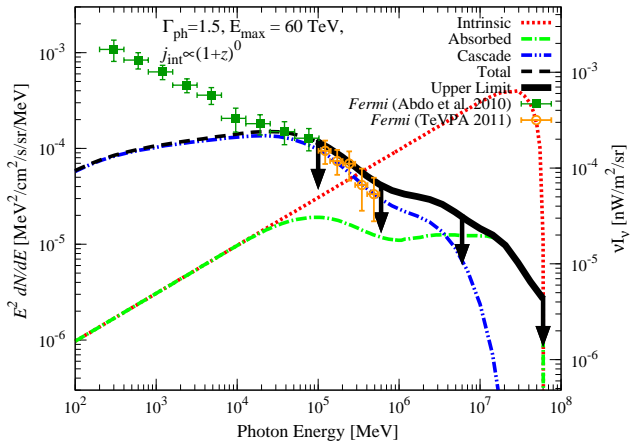}% Here is how to import EPS art
\caption{\label{fig:egb_G1.5} Upper limit on  EGB by requiring the cascade emission not to exceed the EGB data below 100 GeV (cascade-limit case) in the model-independent way. The observation is still consistent with this upper limit. We set the photon index $\Gamma_{\rm ph}=1.5$, the $z$-evolution index $\beta_{\rm evo}=0.0$, and the maximum energy $E_{\rm max}=60.0$ TeV in Eq.~(\ref{eq:jint}). Dotted, dot--dashed, double dot--dashed and dashed curves show the intrinsic spectrum (no absorption), absorbed, cascade, and total (absorbed+cascade) EGB spectrum, respectively. Thick solid curve with arrows show the upper limit.  The filled square points show the observed EGB data with the 11-months {\it Fermi} data \cite{abd10_egrb}. The circle points show the observed EGB data with the 24-months {\it Fermi} data \cite{ack11_TeVPA}. Error bars represent 1-$\sigma$ uncertainty of the data.}
\end{figure}

\section{Results}
\label{sec:res}
There are three main parameters: the photon index $\Gamma_{\rm ph}$; the $z$-evolution index $\beta_{\rm evo}$; and the maximum energy $E_{\rm max}$ in Eq.~(\ref{eq:jint}). As shown later, $\Gamma_{\rm ph}\approx1.5$ gives the most conservative limit on EGB. $\Gamma_{\rm ph}=1.5$ is also expected to be the hardest photon index in the diffusive shock acceleration  \cite{mal01}.

Most of astrophysical sources show positive cosmological evolution ($\beta_{\rm evo}>0$) with the comoving number density increasing with redshift. High-frequency-peaked BL Lacs (HBLs), elliptical galaxies, and cluster of galaxies are known to show no or negative evolution, $\beta_{\rm evo}=0$ \cite{pad07}, $\beta_{\rm evo}=-0.86$ \cite{im02}, and $\beta_{\rm evo}=-1.0$ \cite{mul04},  respectively. Hereafter, we study $\beta_{\rm evo}\le 0$ to minimize the EBL effect, otherwise we notice it.

\begin{figure}[t]
\includegraphics[width=1\linewidth]{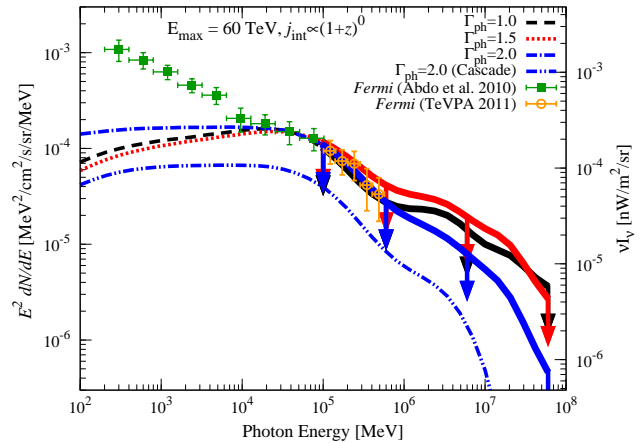}% Here is how to import EPS art
\caption{\label{fig:egb_G2.0} Upper limits on the EGB spectrum for various photon index $\Gamma_{\rm ph}$ parameters are shown here. Dashed, dotted, and dot-dashed curves correspond to $\Gamma_{\rm ph}=$ 1.0, 1.5, and 2.0, respectively. Thick solid curves represents the upper limit for each parameters. For the case of $\Gamma_{\rm ph}=2.0$, we show the cascade component by double dot--dashed curve. The upper limit on EGB with $\Gamma_{\rm ph}=2.0$ is set by requiring the primary component not to exceed the observation (primary-limit case). The upper limits on EGB for $\Gamma_{\rm ph}=1.0$ and 2.0 are tighter than the case of $\Gamma_{\rm ph}=1.5$. The observational data are the same as those in Fig. \ref{fig:egb_G1.5}.}
\end{figure}

\subsection{Model-Independent Self-Limitation method}

Figure \ref{fig:egb_G1.5} shows a typical example of the self-limitation method. Given $(\Gamma_{\rm ph}, \beta_{\rm evo}, E_{\rm max})=(1.5, 0.0,60\ {\rm TeV})$, we can only adjust the normalization of EGB by requiring the cascade emission not to exceed the data observed by {\it Fermi} \cite{abd10_egrb} (cascade-limit case). Then we obtain the upper limit above 100 GeV, well approximated  as
\begin{equation}
E^2\frac{dN}{dE}<1.1\times10^{-4}
\left(\frac{E}{100\,{\rm GeV}}\right)^{-0.5}\  {\rm MeV/cm^2/s/sr}.
\label{eq:limit}
\end{equation}
 This is still consistent with the current observation. The normalization of the EGB upper limit is determined by the observed EGB data at $\sim$10 GeV. Although $\sim$60 TeV emission from extragalactic sources has not been observed yet, we adopt $E_{\rm max}=60$ TeV here to constrain the VHE EGB. We show the cases with different maximum energies and spectral models later. Here we do not include the guaranteed source classes' contribution to EGB. As shown below, the limit violates the observation once we take into account the known source's contribution such as flat spectrum radio quasars (FSRQs), BL Lacs, radio galaxies, and starburst galaxies.

Figure \ref{fig:egb_G2.0} is the same as Figure \ref{fig:egb_G1.5} but for $\Gamma_{\rm ph}=1.0$, $1.5$, and $2.0$. Even if we increase the normalization of the input EGB or take $\Gamma_{\rm ph}<1.5$ to explain the VHE EGB data, the cascade flux increases at the same time.  Then the limit becomes stronger than the case with $\Gamma=1.5$. The model with $\Gamma_{\rm ph}=2.0$ shows a typical example that EGB is limited by the primary component (primary-limit case). Softer input spectrum results in a stronger upper limit. This is because the VHE EGB spectrum is determined by the absorbed component alone, not by the cascade component, although the cascade spectral shape is almost independent of the primary spectrum. Thus, $\Gamma_{\rm ph}\approx1.5$ is the most conservative case.

\begin{figure}[t]
\includegraphics[width=1\linewidth]{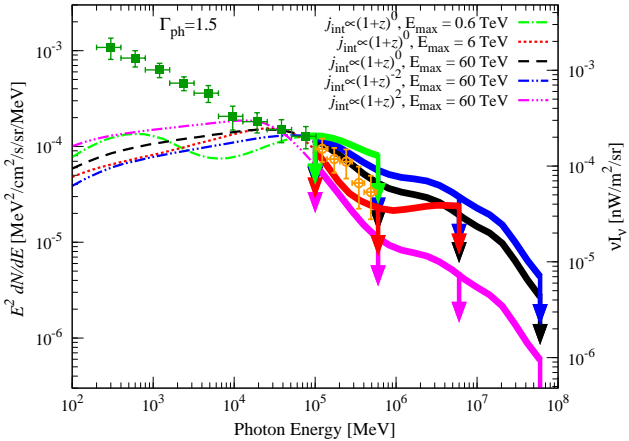}% Here is how to import EPS art
\caption{\label{fig:egb_G1.5_eb} Upper limits on the EGB spectrum for various parameters are shown here. We show the total (absorbed + cascade) spectrum with the photon index $\Gamma_{\rm ph}=1.5$. Dot--dashed, dotted, and dashed  curves correspond to the maximum energy $E_{\rm max}$ = 0.6, 6.0, 60 TeV for $\beta_{\rm evo}=0.0$, respectively. Double dot--dashed, and triple dot-dashed curves are the limits for $\beta_{\rm evo}$=-2.0, and 2.0 for $E_{\rm max}$ = 60 TeV, respectively. Thick solid curves with arrows represent the upper limits for each parameter sets. The observational data are the same as those in Figure \ref{fig:egb_G1.5}. }
\end{figure}

Figure \ref{fig:egb_G1.5_eb} shows the upper limits on EGB for $\Gamma_{\rm ph}=$ 1.5 with various $\beta_{\rm evo}$ and $E_{\rm max}$. Negative evolution with $\beta_{\rm evo}<0$ eases the upper limit on EGB. In the negative evolution case, the dominant EGB contribution comes from low redshift and the absorption effects are small. Since a large fraction of VHE emitting sources locates inside the gamma-ray horizon without suffering the EBL attenuation, the contribution of cascade emission becomes minor. The maximum energy as low as $E_{\rm max}=0.6$ TeV also eases the upper limit. This is because the cascade emission appears only at $\sim0.3$ GeV following Eq. \ref{eq:ecas}. However, there are no known sources that have a large contribution to EGB and a spectral cutoff at $\sim$TeV. For example, imaging atmospheric Cherenkov telescopes (IACTs) detects $>0.6$ TeV emission from several nearby blazars \cite[e.g.][]{abd11_mrk421}.

Each curve in Fig. \ref{fig:egb_G1.5_eb} gives the EGB upper limit in each energy band for a fixed $\beta_{\rm evo}$ because the limit is basically set by the original flux before
absorption at the maximum energy. For $\beta_{\rm evo}=0$, the limit is approximated by Eq.~(\ref{eq:limit}). Note that $\Gamma_{\rm ph} \approx 1.5$ is the most conservative case as discussed above.

With $\beta_{\rm evo}=2.0$, i.e. positive evolution, the upper limit violates the EGB measurement for $(\Gamma_{\rm ph},  E_{\rm max})\ = \ (1.5,\ 60 \ {\rm TeV})$. To avoid the inconsistency with the measured EGB, the VHE emissivity beyond the gamma-ray horizon should be turned off or less than that inside the horizon. Therefore, if the EGB origin is cosmological, the source might have hard spectra and show a no or negative cosmological evolution.  Recent {\it Fermi} analysis shows that FSRQs have $\beta=5.7$ \cite{aje12}. No or negative-evolution sources reported in gamma-ray or in other wavelength are HBLs \cite{pad07}, elliptical galaxies \cite{im02} and clusters of galaxies \cite{mul04}. Gamma-ray emission from latter two has not been confirmed yet and is not likely enough for EGB at least in the case of cluster of galaxies \cite{gab03}.

\begin{figure}[t]
\includegraphics[width=1\linewidth]{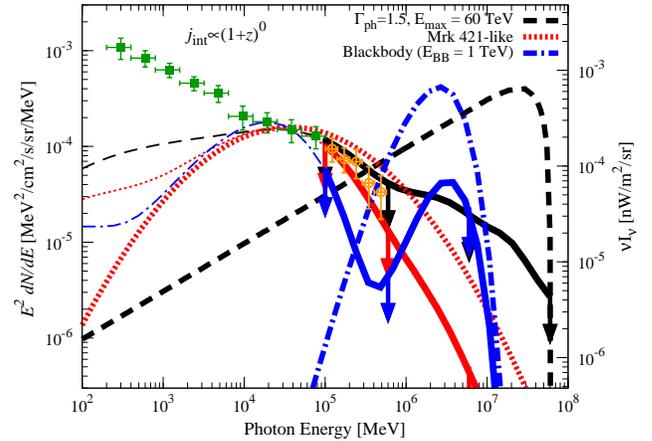}% Here is how to import EPS art
\caption{\label{fig:egb_type} Upper limits on the EGB spectrum for various spectral models with $\beta_{\rm evo}=0.0$ are shown here. Dashed, dotted, and dot-dashed curves correspond to ($\Gamma_{\rm ph}, E_{\rm max})=(1.5, 60 \ {\rm TeV})$, Mrk 421-like spectrum, and Blackbody spectrum with $E_{\rm BB}=1$ TeV, respectively. Thick and thin curve represents the intrinsic spectrum and the total (absorbed+cascade) spectrum, respectively. Thick solid curves represent the upper limit for each parameter. The observational data are the same as those in Fig. \ref{fig:egb_G1.5}.}
\end{figure}

Fig. \ref{fig:egb_type} shows the upper limits for non-power-law spectral models in the case of no evolution. Here we show a Mrk 421-like spectrum and a blackbody shape spectrum. For a Mrk 421-like spectrum, we use a log-parabola function as 
\begin{equation}
\frac{dj_{\rm int}}{dE_\gamma}(E_\gamma,z)\propto \left(\frac{E_\gamma}{E_{\rm br}}\right)^{-\Gamma_{\rm ph}+\delta\log(\frac{E_\gamma}{E_{\rm br}})}\times(1+z)^{\beta_{\rm evo}},
\end{equation}
where we use the best fit parameters for Mrk 421 as $E_{\rm br} = 0.3$ TeV, $\Gamma_{\rm ph}$ = 2.48, and $\delta=0.33$ \cite{abd11_mrk421}. For a blackbody shape spectrum, we adopt 
 \begin{equation}
\frac{dj_{\rm int}}{dE_\gamma}(E_\gamma,z)\propto \frac{E_\gamma^2}{\exp(E_\gamma/E_{\rm BB})-1}\times(1+z)^{\beta_{\rm evo}},
\end{equation}
where we set $E_{\rm BB}=1$ TeV. The upper limits for both of a Mrk 421-like spectrum and a blackbody spectrum models comes lower than that for $\Gamma=1.5$ spectral model. Therefore, $\Gamma\approx1.5$ is the most conservative case even if we consider these non-power-law spectral models.

\begin{figure}[t]
\includegraphics[width=1\linewidth]{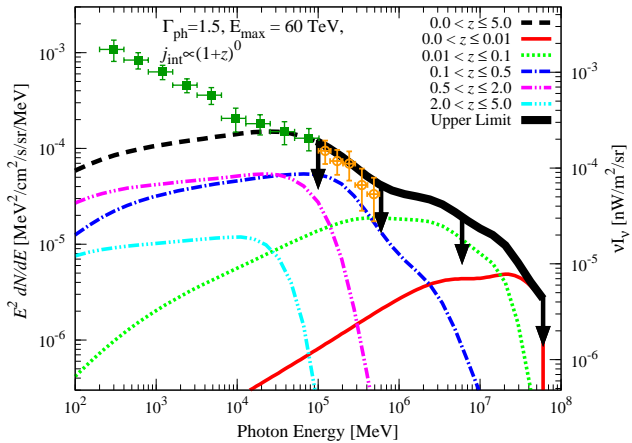}% Here is how to import EPS art
\caption{\label{fig:egb_z} Upper limits on the EGB spectrum for various redshift bins are shown. Thick solid curve with arrows shows the upper limit. Solid, dotted, dot--dashed, double dot--dashed, triple dot-dashed and dashed curves show $0<z\le0.01$, $0.01<z\le0.1$, $0.1<z\le0.5$, $0.5<z\le2$, $2<z\le5$, and $0<z\le5$ EGB spectrum, respectively. }
\end{figure}

Fig. \ref{fig:egb_z} shows the upper limit for $(\Gamma_{\rm ph}, \beta_{\rm evo}, E_{\rm max})=(1.5, 0.0,60\ {\rm TeV})$. Here we show the contribution from each redshift ranges. At VHE region, only sources at $z<0.5$ can contribute to the EGB due to the EBL suppression. This means that the dominant VHE EGB contribution at each energy roughly comes from inside the gamma-ray horizon by EBL attenuation. Therefore, we can constrain the VHE emissivity of the universe at each energy by future VHE EGB measurements.

\subsection{Self-Limitation method with known sources' contribution}
There are guaranteed source classes that contribute to EGB detected by EGRET or {\it Fermi}. It is expected that blazars, radio galaxies, and starburst galaxies explain $22.5\pm1.8$\% \cite{abd10_marco}, $25^{+38}_{-15}$\% \cite{ino11} and $4-23$\% \cite{ack12_stb} of the unresolved EGB, respectively. Then, $\sim70$\% of EGB will be explained by known source classes. We need to subtract them to evaluate the VHE EGB upper limit, since the residual is the only room for the cascade plus absorbed emission from VHE EGB.

For FSRQs, we adopt the model by \cite{aje12} (hereafter MA12). For BL Lacs, we use the model by \cite{abd10_marco} (hereafter Fermi10). Since the EBL absorption effect is not taken into account in \cite{abd10_marco}, we include the EBL attenuation model \cite{fin10} by assuming $\beta_{\rm evo}=0$ for BL Lacs. For radio galaxies, we use the model by \cite{ino11} (hereafter YI11). For starburst galaxies, we use the IR luminosity function model from \cite{ste11}  (hereafter SV11) and the power law model from \cite{ack12_stb} (hereafter Fermi12). We renormalize the SV11 model by a factor of 0.8 to avoid the total (FSRQ+BL Lacs+radio galaxy+starburst galaxy) contribution exceeding the observed EGB data, or the VHE upper limit becomes zero.

\begin{figure}[t]
\includegraphics[width=1\linewidth]{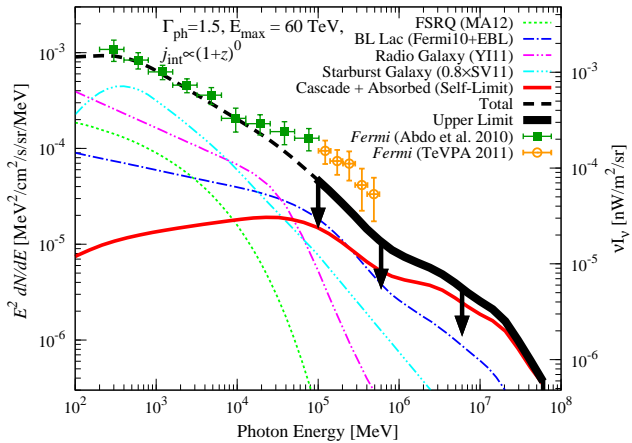}% Here is how to import EPS art
\caption{\label{fig:egb_bl12} 
 Upper limit on the EGB spectrum by adding the known sources' contribution for $(\Gamma_{\rm ph}, \beta_{\rm evo}, E_{\rm max})=(1.5, 0.0,60\ {\rm TeV})$. The dotted, dot-dashed, double dot-dashed, and triple dot-dashed curves correspond to contribution from FSRQ \cite[MA12]{aje12}, BL Lac \cite[Fermi10]{abd10_marco}, radio galaxy \cite[YI11]{ino11}, and starburst galaxy \cite[SV11]{ste11}, respectively. For BL Lac, we include EBL absorption effect \cite{fin10} by assuming $\beta_{\rm evo}=0$. For starburst galaxy, we use the SV11 IR luminosity function model renormalized by a factor of 0.8. The thin solid curve show the cascade+absorbed contribution from VHE EGB (self-limit method). The dashed curve show the total emission of FSRQs, BL Lacs, radio galaxy, starburst galaxy and our model. The thick solid curve with arrows shows an upper limit which is approximated by Eq. \ref{eq:limit_mo}. The observational data are the same as those in Figure \ref{fig:egb_G1.5}. }
\end{figure}

Figs. \ref{fig:egb_bl12} and \ref{fig:egb_fermi12} shows the upper limit on the EGB taking into account the known sources' contributions where we use the SV11 model and the Fermi12 model for starburst galaxies, respectively. Here we show the case of $(\Gamma_{\rm ph}, \beta_{\rm evo}, E_{\rm max})=(1.5, 0.0,60\ {\rm TeV})$. The upper limit on EGB is derived from the sum of the VHE EGB cascade and guaranteed sources' contribution. When we try to explain the EGB below 10 GeV by guaranteed sources as in Fig. \ref{fig:egb_bl12}, the EGB measurement violates the upper limit above 100 GeV. The limit is approximated by Eq. \ref{eq:limit_mo}. If we take $\beta_{\rm evo}=-4$, the upper limit becomes consistent with the measured spectrum with 1-sigma difference. However, there are no known sources showing such a strongly negative evolution in any wavelengths. In the case of $E_{\rm max}=6$ TeV, $\beta_{\rm evo}<-6$ is required. Therefore, the VHE emissivity at redshift $z\gtrsim0.5$ should be low. On the other hand, when we try to make the upper limit consistent with the VHE EGB data as in Fig. \ref{fig:egb_fermi12}, the total EGB contribution from cosmological sources is $\sim$2 sigma below the measured EGB below 10 GeV. In this case, the limit is approximated by 1.6 times Eq. \ref{eq:limit_mo}.

\section{Discussion and Conclusion}
\label{sec:dis}

\begin{figure}[t]
\includegraphics[width=1\linewidth]{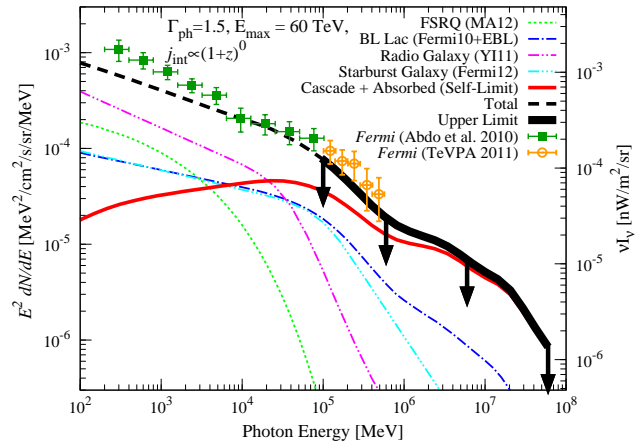}% Here is how to import EPS art
\caption{\label{fig:egb_fermi12} Same as Fig. \ref{fig:egb_bl12}, but we use the Milky Way model by \cite[Fermi12]{ack12_stb} for starburst galaxies. The thick solid curve with arrows shows an upper limit which is approximated by 1.6 times Eq. \ref{eq:limit_mo}.}
\end{figure}

There are a few possible scenarios to explain VHE EGB.  (i) Sources have hard spectra $\Gamma_{\rm ph} \approx 1.5$ with a cutoff at 60 TeV and a strongly negative evolution $\beta_{\rm evo} \lesssim -4$ in Eq.~(\ref{eq:jint}). If a cutoff is at 6 TeV, $\beta_{\rm evo}\lesssim-6$ is required. (ii) More transparent EBL generates weaker absorption effect and eases the limit. However we use the EBL model close to the minimum (integrated flux of galaxies). Even if we use it, the upper limit is still below the observation.  (iii) Pair production process could be affected by new physics such as Lorentz-invariance violation \cite{jac08} and axion-like particles \cite{hor12}. (iv) Dark matter annihilation/decay in the local group \cite{zel80,pie04} can avoid the EBL absorption effect. (v) Sources only contributing to EGB  at $\le10$ GeV may complement the residual between data and model in Fig. \ref{fig:egb_fermi12}. High latitude pulsars and radio-quiet AGNs are possible candidates. First, pulsars observed in gamma-ray have a cutoff at $\sim5$ GeV \cite{abd10_pulsar}. Second, radio-quiet AGNs may contribute to the EGB at $\le10$ GeV \cite{ino08}, although {\it Fermi} does not see radio-quiet AGNs \cite{ten11,ack12_seyfert}. If non-thermal electrons exist in a corona above the accretion disk, a power-law tail will appear in hard X-ray and gamma-ray band \cite{ino08}. (vi) The EGB measurement has uncertainties. The EGB is deduced by subtracting the foreground emission from our Galaxy which is still not fully understood. For example, in the analysis in \cite{abd10_egrb}, the {\it Fermi} bubble \cite{su10} is not subtracted. 

In the scenario (i), we can reject known {\it Fermi} gamma-ray source classes as the origin of VHE EGB. First, blazars and radio galaxies detected by {\it Fermi} do not show negative cosmic evolution \cite{abd10_marco,ino11}. Second, gamma-ray observed galaxies do not show $\Gamma_{\rm ph}=1.5$ \cite{abd10_gal}. Even if they can create such hard spectra, TeV emission is internally absorbed by the interstellar radiation \cite{ino11_stb}. Third, pulsars will not contribute at VHE band as discussed above. Although a power-law tail in the VHE band have been recently reported for the Crab pulsar \cite{ali11}, the photon index $\Gamma_{\rm ph}=3.8$ is softer than that of the observed VHE EGB $\Gamma_{\rm ph}=2.41$.

One of the most likely source classes is TeV selected HBL (TeV HBL) which is not detected by {\it Fermi} but by the current IACTs \cite{abd09_TeV}. Although their cosmological evolution is still unknown, their spectrum is hard and could have a cut-off at $\sim10$ TeV. Interestingly these hard TeV emission can be explained by high energy cosmic-ray induced intergalactic cascade \cite{mur11}. A new IACT array Cherenkov Telescope Array (CTA) \cite{act11} is expected to detect $>100$ blazars including TeV HBLs \cite{ino10a}. CTA will  enable us to statistically study their evolution and contribution to VHE EGB.

Low luminosity (LL) GRBs may also explain VHE EGB, although LL GRBs have not been detected in gamma-ray. LL GRBs might show a negative cosmological evolution since LL GRBs are only discovered at low redshift \cite[e.g.][]{cam06}. The total energy budget of LL GRBs is large enough to explain Ultra-High Energy Cosmic Rays (UHECRs) \cite{mur06}. The UHECR intensity is $\sim1.0\times10^{-5} \  {\rm MeV/cm^2/s/sr}$ at $\sim10^{19}$ eV \cite{kot11} which is comparable to the observed EGB intensity $\sim3.0\times10^{-5} \  {\rm MeV/cm^2/s/sr}$ at 600 GeV \cite{ack11_TeVPA}.

It is important to measure VHE EGB more precisely and at higher energy, such as by {\it Fermi}, CTA and CALET (CALorimetric Electron Telescope) \cite{tor08}. Our upper limit above 600GeV will help future measurements to unveil the EGB origin. Just by detecting the EGB above TeV, we can put a meaningful lower limit on the number density of the EGB sources because a source should
reside in the gamma-ray horizon that is small at high energy. For
example, if CTA measures the EGB at 60TeV, the gamma-ray horizon is $\sim40$ Mpc and the EGB source number in the entire sky should be larger than $N_{\rm min} = 4\pi F_{\rm EGB}(60 {\rm TeV})/F_{\rm CTA}(60 {\rm TeV}) \sim 6[F_{\rm EGB}(60 {\rm TeV})/3\times10^{-7} \ {\rm MeV/cm^2/s/sr}]$ \cite[see also][]{mur12,mur12c},
where we assume CTA sensitivity as $F_{\rm CTA}(60 {\rm TeV})=1.0\times10^{-12} {\rm erg/cm^2/s}$ \cite{act11}. We note that the total emissivity within the gamma-ray horizon may have a large dispersion (such as Poisson fluctuation) at high energy because of the small horizon size, which may lead to a violation of our upper limits. In other words, the number of sources within the gamma-ray horizon may be larger at high energy region than our expectation due to the local distribution fluctuation.

Anisotropy in the VHE EGB above 30 GeV is an alternative key to understand VHE EGB, since the monopole peak depends on sources \cite{and07}. The anisotropy below 50 GeV has already been investigated \cite{ack12_aniso}.

The effect of IGMF to the cascade component is not critical in this study. However, with weak strength of IGMF, {\it Fermi} may have already detected the cascade emission alone from some sources and classified them into unassociated sources \cite{yan12}. If their contribution had been already subtracted from the EGB, the upper limit on the VHE EGB would be eased. If the cascade emission alone had been detected, the absorbed VHE flux would be brighter than the {\it Fermi}'s $\nu F_\nu$  sensitivity. Then, the current IACTs should have detected those VHE emission, since their sensitivity in $\nu F_\nu$ in the VHE region is comparable to that of {\it Fermi} in the GeV region. However, TeV HBLs have not been detected by {\it Fermi} yet \cite{abd09_TeV}. Therefore, there would be no sources whose cascade component alone is resolved.

Strong IGMF would also ease the upper limit on the VHE EGB. When the IGMF strength is above 3.26 $\mu$G, the magnetic energy density takes over the CMB energy density in the Thomson regime. Then, the synchrotron cooling is more effective than the IC cooling. The strong IGMF will suppress the cascade emissivity in gamma-ray. It is, however, also known that the magnetic field strength of lobes of AGN is typically $1 \mu$G \cite{mas11}. Thus, the strength of IGMF may be smaller than 1$\mu$G with the scale of AGNs' lobes $\sim100{\rm kpc}$ \cite{mas11}. This scale is shorter than the mean free path of pair creation  $\sim 20\ {\rm Mpc}(n_{\rm EBL}(z)/0.1 {\rm cm^{-3}})$. Therefore, strong IGMF would not affect the cascade emissivity in gamma-ray significantly.

The upper limit is also applied to the EGB produced by the UHECRs via the intergalactic cascade \cite[see e.g.][]{ahl10}. In order to explain the VHE EGB by the UHECR cascade with known sources' contribution to EGB, a negative evolution may be necessary for the UHECR sources.

In this paper, we develop a new method to constrain the cosmological EGB by using the EGB itself. VHE photons propagating the universe are absorbed by EBL and create electron--positron pairs. Created pairs generate secondary gamma-ray emission via the IC scattering of the CMB photons. We constrain the VHE EGB by comparing this regenerated emission with the current EGB measurement. Our method also provides upper limits on EGB above 600GeV for future observations, such as {\it Fermi}, CTA and CALET. Our self-limits are useful for identifying the origin of EGB.

We show that the current EGB measurement sets an upper limit of $E^2dN/dE<4.5\times10^{-5}(E/100\ {\rm GeV})^{-0.7}\  {\rm MeV/cm^2/s/sr}$ to the VHE EGB above 100 GeV from cosmological sources, where we take into account the known sources' contributions. The current EGB measurement by {\it Fermi} \cite{ack11_TeVPA} violates the predicted upper limit. In order to make consistent with the observed EGB data in the cosmological origin scenario, possible origins should show strongly negative cosmological evolution, hard spectrum, and a cut-off at $\sim10$ TeV. This kind of sources, however, has never been reported yet  neither in gamma-ray nor in other wavelength.

\begin{acknowledgments}
We thank the anonymous referee for useful comments and suggestions. We thank Marco Ajello and Keith Bechtol for useful discussions and providing their models. We also thank Kohta Murase and Hajime Takami for useful discussions. YI acknowledges support by the Research Fellowship of the Japan Society for the Promotion of Science (JSPS),
and KI
%This work is supported
%in part
by 
%KAKENHI,
the Grant-in-Aid from the 
Ministry of Education, Culture, Sports, Science and Technology
(MEXT) of Japan, 
%Nos. 
21684014, 22244019, 22244030.
% (KI).
\end{acknowledgments}
\providecommand{\noopsort}[1]{}\providecommand{\singleletter}[1]{#1}%

\end{document}